\documentclass[twocolumn,showpacs,preprintnumbers,amsmath,amssymb,prl]{revtex4}

\usepackage{graphicx}
\usepackage{bm}

\begin{document}

\title{Order parameter oscillations in Fe/Ag/Bi$_2$Sr$_2$CaCu$_2$O$_{8+\delta}$ tunnel junctions}

\author{Mario Freamat and K.-W. Ng}
\affiliation{Department of Physics and Astronomy, University of Kentucky, Lexington,
KY 40506-0055, U.S.A.}

\date{\today}

\begin{abstract}
We have performed temperature dependent tunneling conductance spectroscopy on
Fe/Ag/Bi$_2$Sr$_2$CaCu$_2$O$_{8+\delta}$(BSCCO) planar junctions. The
multilayered Fe counterelectrode was designed to probe the proximity region of
the ab-plane of BSCCO. The spectra manifested a coherent oscillatory behavior
with magnitude and sign dependent on the energy, decaying with increasing distance
from the junction barrier, in conjunction with the theoretical predictions
involving d-wave superconductors coupled with ferromagnets. The conductance
oscillates in antiphase at E = 0 and E = $\pm\Delta$. Spectral features
characteristic to a broken time-reversal pairing symmetry are detected and they do
not depend on the geometrical characteristics of the ferromagnetic film.
\end{abstract}

\maketitle

In recent years numerous theoretical and experimental studies have been devoted
to the spin polarized transport in ferromagnet(FM)-superconductor(SC) structures.
This interest is equally motivated by the rapidly developing applications, e.g.
in quantum two-level systems (qubits) based on phase shifts \cite{Orl}, as well as
by the very interesting physics involving unconventional superconducting states.
The exchange energy $h$ in the ferromagnet leads to a spatially inhomogeneous
state with the order parameter presenting phase changes by $\pi$ in the proximity
region at the interface with the SC. Such a modulated state is called
Larkin-Ovchinnikov-Fulde-Ferrell (LOFF) from the names of the authors who first
predicted it \cite{Ful}. In the FM, the Cooper paired electrons have different
energies and Fermi momenta, so they dephase on a scale of several ferromagnetic
coherence lengths. Consequently, this region presents a diversity of intensely
explored phenomena like spatially dumped oscillations of the density of states
(DOS) and gapless superconductivity \cite{Buz}-\cite{Sun}, oscillations of the
superconducting critical temperature T$_c$ and critical current I$_c$ with the
thickness of the FM electrode in layered systems \cite{Bou,Rya}, or asymmetric
zero-bias conductance peak (ZBCP) splitting \cite{Saw}.

An extensive attention was dedicated to the LOFF state arising in junctions
between conventional SC's and FM films of variable thickness. As example,
the measurements on Al/Al$_2$O$_3$/PdNi/Nb junctions reported by the Orsay
group \cite{Bou} clearly show the difference in the conductance spectra at two
different locations in the proximity region, indicating the order parameter change
from positive (0-state) to negative values ($\pi$-state). However, the data
originated from experiments involving high-Tc superconductors (HTSC) is rather
scarce, even if it could be conclusive about the interaction between
ferromagnetism and a more complex form of superconductivity. In particular, the
superconductivity mechanism in the cuprates is far from being fully understood.
According to a series of phase sensitive experiments \cite{Tsu} the symmetry of the
order parameter is d-wave, pure or mixed with a secondary s or d component \cite{Cov}.
It also seems that the antiferromagnetic spin plays an important role in the pairing
mechanism \cite{Pin}. Therefore, it is constructive to consider that the LOFF state
in the cuprates, while intrinsically interesting, may also offer information about
what makes the high-Tc superconductivity different from the conventional one.
A multitude of theoretical models were developed to calculate the specific transport
properties, spin and charge relaxation in tunneling experiments involving d-wave
HTSC with FM counterelectrodes. The most popular ones are those based on the
Blonder-Tinkham-Klapwijk (BTK) approach \cite{Kas,Ste}. However, these models
consider just the transmission and reflection processes at the interface with
the FM film modeled as a semi-infinite domain. In this paper we will use a model
\cite{Zar} based on a ballistic quasiclassical formalism with the DOS averaged over
all classical trajectories of multiple reflections in the FM domain of finite
thickness $d$.

Hoping to contribute to the experimental information about this topic, we report in
this paper the transport properties in a
multilayered-Fe/Ag/Bi$_2$Sr$_2$CaCu$_2$O$_{8-\delta}$ (BSCCO) ab-plane
oriented planar tunneling junction. The BSCCO ab-plane is probed successively
from 5 overlapping Fe thin films with  the same thickness, through an intermediate
thin layer of Ag designed to minimize the barrier strength and avoid the appearance
of a spin-glass phase at the junction interface \cite{Rub}. The spectra are
collected on the thicker and thicker counterelectrodes and studied as function
of the temperature and thickness of the FM domain.
\begin{figure}
\includegraphics{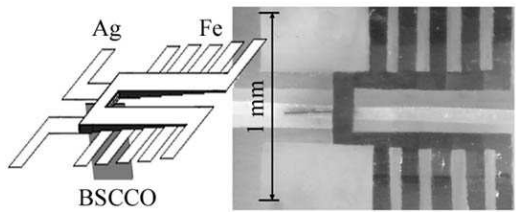}
\caption{\label{fig:F1} Junction layout. The layered Fe/Ag conterelectrode
is deposited perpendicular on the BSCCO ab-plane.}
\end{figure}
The planar junctions were built on freshly cleaved monocrystal samples of slightly
underdoped BSCCO grown by the self-flux method. Two silver paste leads were
attached to the crystal and its lower side was molded in epoxy resin. Before
the epoxy dried, the upper side was fixed between two quartz glass slabs
with coplanar surfaces. The epoxy penetrated between the sides of the slabs
so that the crystal was caught between two smooth glass surfaces attached by
thin epoxy walls. While the surface of the glass was protected, we performed a
clean mechanical polishing along the ab-plane of the crystal, laterally
embedded in epoxy. We monitored the process with a profilometer, until the
exposed edge of the crystal was at the same level as the glass surfaces.
The counterelectrode sandwich was then evaporated using a metallic shadow
mask shaped as in Fig.~\ref{fig:F1}, forming thin film strips 0.1 mm in width.
The deposition was controlled using a calibrated quartz thickness-monitor.
Above a 30\AA\ silver film, five 30\AA\ iron films were deposited at a rate
of 0.1\AA/s for uniformity \cite{Raz}. The residual resistivity ratio of the
Fe film $RRR = R(300K)/R(40K)\approx 8.3$ indicates a good deposition quality.
The tunneling barrier is formed naturally on the exposed edge of BSCCO. It
had a resistance $R_B(300K) \approx 450 \Omega$, for the typical junction
presented here, showing a high transparency. We evaluated the elastic mean
free path $l$ of conduction electrons in one of our Fe films using an expression
derived from Pippard relations \cite{Raz}:
$v_Fl=\left(\pi k_B/e\right)^2\left(\sigma/\gamma\right)$.
With the iron film conductance $\sigma\approx$4$\times$10$^3$ S/m$^2$,
electronic specific heat $\gamma$= 5$\times$10$^{-3}$ J/K$^2$, and Fermi
velocity $v_F$ = 1.98$\times$10$^6$m/s, we obtained $l\approx$100\AA. Since the
quasiclassical description of conductance assumes $l<d$ \cite{Zar},
we see that such a model is applicable to our first three Fe layers.
We performed conventional 4-point measurements, with a constant current
driving bias.

\begin{figure}
\includegraphics{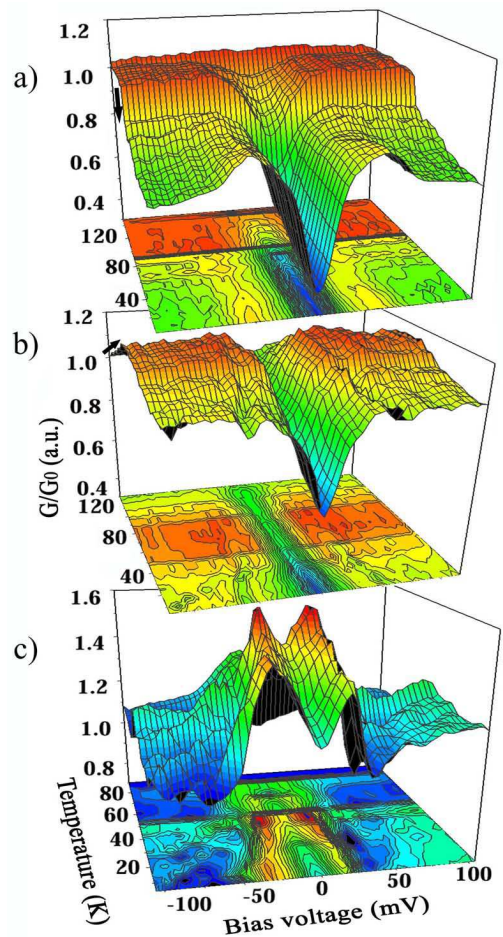}
\caption{\label{fig:F2} Temperature dependence of the conductance spectra. The arrows
indicate the background conductance variation at $T_c$ as the temperature is decreased.
a) Normal layer b) 30\AA\ iron layer, and c) 60\AA\ iron layer.}
\end{figure}

\begin{figure*}
    \begin{minipage}[c]{0.51\linewidth}
      \includegraphics{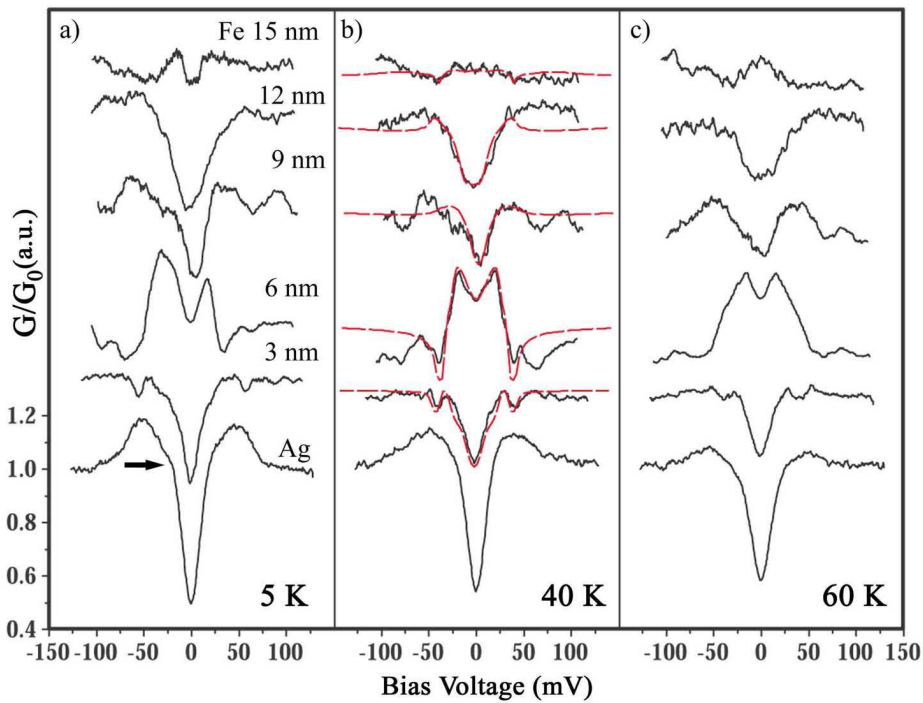}
    \end{minipage}\hfill
    \begin{minipage}[c]{0.29\linewidth}
      \caption{\label{fig:F3} Tunneling spectra normalized by the background conductance
      $G_0$, represented for increasing counterelectrode thickness $d$, at three
      temperatures $T$. a) $T$ = 5 K. The arrow on the Ag/BSCCO spectrum points to the
      minigap edge. The asymmetric spectrum at $d$ = 60\AA\ permits an evaluation of
      the Fe film polarization from the peak ratio, $P_{Fe}\approx 43\%$.
      b) $T$ = 40 K spectra (continuous lines) compared to the theoretical fits (dash lines)
      with $h$ = 170 meV (20 meV for the first Fe layer), $\Delta =$ 44 meV, and Z = 0.51.
      The fitted $d$ values (Fig.~\ref{fig:F2}b) match close enough the actual values.
      c) $T$ = 60 K. The ZBCP for $d$ = 150\AA\ is not split at this higher
      temperature approaching $T_c$, due to the suppression of the BTRS and the
      bigger effect from the spin-splitting field on the superconductivity
      since $\Delta$ decreases while the $h$ remains constant.}
    \end{minipage}
\end{figure*}

The tunneling spectra at the interface with the BSCCO crystal were obtained
successively for the six layers, starting with the Ag counterelectrode, then
probing an increasing thickness in the FM region from 0 to 150\AA, by 30\AA
increment for each electrode. Each layer was measured at different temperatures,
from 150 to 4.2 K. Fig.2 represents the temperature dependence for the first
three layers, each curve being normalized by the spectrum at 150K. All layers
manifested a sudden change in the background conductance at T$_c$. This phenomenon
can be attributed to the competition between a screening spin-accumulation close
to the interface, responsible for a drop in the barrier conductance, and the onset
of Andreev reflections, with an adverse effect \cite{Bel}. In an Andreev
reflection, the electron incident on the barrier with energy lower than the
superconductive gap is reflected as a hole, while a Cooper pair is transmitted
into the SC. In the case of d-wave superconductors the Andreev reflections are
phase sensitive. This leads to the formation of bound states at the junction
interface with amplitudes dependent on the angle of incidence. In a quasiclassical
picture \cite{Zar}, the particles move along trajectories reflected between the
boundaries of the counterelectrode, the density of states per trajectory
being correlated to the length of the trajectory or number of reflections
which in turn depends on the energy and exchange potential. The tunneling
conductance spectrum is a measure of the energy resolved total density of states
and is regarded as a sum of local bound states averaged over all trajectories
distributed by length. Therefore, the thickness of the counterelectrode influences
the amplitude of the bound states and thus the spectral features, like the
ZBCP. For our Ag terminal, since the amplitude of the Andreev states is
low (Fig.~\ref{fig:F2} and ~\ref{fig:F3}), and it is in direct contact with
the next iron layer, the background conductance at $T_c$ decreases with
decreasing temperature, due to the predominance of a spin accumulation at
the junction barrier. The temperature dependence of the Ag/BSCCO tunneling
spectrum shows no ZBCP (Fig.~\ref{fig:F2}a). This may indicate a predominant
injection normal on the [100] plane \cite{Kas,Ste} combined with a rarefied
distribution of long trajectories due to the very low thickness of the film,
leading to the suppression of low energy states. The energy resolved DOS shows
signs of a minigap (Fig.~\ref{fig:F3}) at zero energy bias, which points out to
some degree of disorder in the counterelectrode favorable to short trajectories or
higher energy subgap bound states. A pseudogap opens at $T^*\approx$ 160 K and the
superconducting transition takes place at $T_c \approx$ 85 K, clearly visible on
the T-V plane contour. We measured a peak to peak SC energy gap
$\Delta \approx$ 44 meV at 40 K.

\begin{figure}
\includegraphics{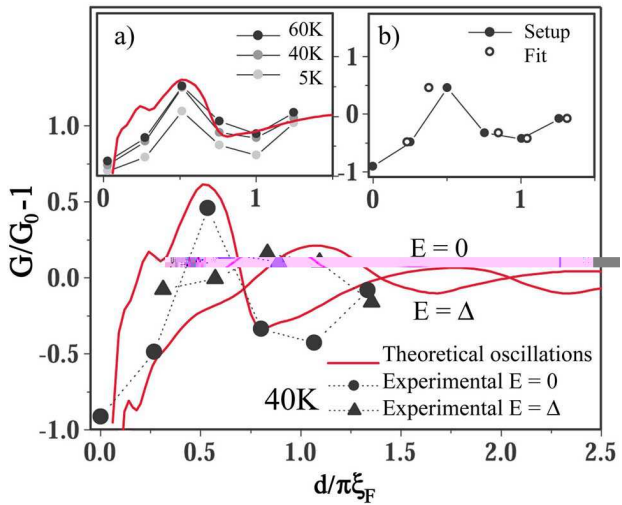}
\caption{\label{fig:F4} The reduced DOS develops coherent oscillations
dependent on $d$. a) The experimental results approach better the predicted
behavior for increasing temperature, due to the decreased s-wave component.
b) There is a remarkable match between the theoretical thicknesses
and the actual $d$ values.}
\end{figure}

Next we probed the FM region of the junction. With each Fe layer,
30\AA\ are added to the total distance accessible to the pairs injected into
the counterelectrode. The pairs survive along a distance of several coherence
lengths $\xi \approx$ 38\AA\ as the SC order parameter decays oscillatory around
zero. We performed theoretical fits for the DOS spectra at the intermediate
temperature 40 K. The model we employed \cite{Zar} calculates the DOS taking
into account the thickness $d$ of the FM film when the interface with the SC is
highly transparent and $d$ is smaller than the electron mean free path $l$.
With $l$ calculated above, the model is applicable mostly for the first few layers,
which are anyway the most suggestive. Since we expect our boundaries to be rough,
leading to an admixture of trajectories and a smearing of the spectral features,
the DOS was averaged over a Gaussian distribution of thicknesses around each
$d$. The best fits imposed a value $h$ = 20 meV for the first iron layer and
$h$ = 170 meV for the second one (consistent with the accepted value \cite{Raz}).
The fits for all other Fe layers were obtained by keeping constant the exchange
energy  $h$ = 170 meV, the gap $\Delta =$ 44 meV, and the barrier transparency
Z = 0.51. The only adjusted variable was the thickness scaled by the coherence
length, $d/\pi\xi_F$. The procedure resulted into five values for the layer
thickness remarkably close to the actual values (see Fig.~\ref{fig:F4}b),
confirming the validity of the model and also the regularity of the thin films.
We also take advantage from the BTK type models to underline the effect of the
d-wave (possibly mixed with s-wave) pairing symmetry of BSCCO.

The temperature dependent spectra measured on the first FM layer ($d$ = 30\AA)
(Fig.~\ref{fig:F2}b) also has a pseudogap. The $T_c$ is slightly smaller
and the background conductance after the transition increases, indicating
the growing influence of the Andreev reflections. The most remarkable
features (Fig.~\ref{fig:F3}), are the dips replacing the $\pm\Delta$
coherence peaks. Due to the direct contact with the Ag layer, the spectrum
rather shapes up the effect of a distribution of spin-polarized quasiparticles
in the Ag film, so that the theoretical fit on this curve imposed an exchange
energy $h$=20 meV, lower than that for the other Fe layers. For 0$<h<\Delta$,
the $\pm\Delta$ peaks move to subgap energies, being replaced by minima in the
DOS \cite{Zar}. The resulted subgap peaks are expected to merge into a ZBCP
with increasing $d$, and to exhibit local oscillatory variations. Indeed,
the next FM layers show the oscillations, but the peaks do not merge with
increasing $d$. This may be due to properties of the junction unaffected by
the thickness of the FM environment, like the presence of a mixed superconductive
phase with broken time-reversal  symmetry (BTRS) at the junction interface
(e.g., (d+is)-wave)\cite{Cov}. This splits the ZBCP, forming a gaplike feature
with the size given by the magnitude of the sub-component s, and influenced by
temperature. In this case, the s-wave gap remains unchanged with increasing $d$,
while the subgap DOS oscillates up and down as the decaying order parameter
alternatively takes negative and positive values respectively.

As seen in Fig.~\ref{fig:F2}c, the next Fe layer ($d$ = 60\AA) measures a
flatter pseudogap at high temperatures. The conductance enhancement at the
superconductive transition is larger, consistent with the high amplitude of
the Andreev peaks. The electronlike and holelike quasiparticle branches of
the spectrum are asymmetric at lower temperatures. The split ZBCP is high
and it becomes broader and less asymmetric with increasing temperature
(Fig.~\ref{fig:F3}a,c). This asymmetry may be attributed \cite{Kas} to the
presence of a spin dependent contribution to the ZBCP split at lower
temperatures. If a spin imbalance is present in the barrier, the ZBCP
corresponding to up (down) spin currents are shifted at higher (lower)
energies. The polarization in the FM layer decreases the spin-up peak and
increases the spin-down peak. The ratio of the peaks can be used to evaluate
the spin polarization $P$ in iron. In our case, $P_{Fe}\approx$ 43\%,
which is consistent with the values reported in literature \cite{Raz}.
The broader ZBCP at temperatures closer to $T_c$ may be the effect of a
thermal smearing and the suppression of superconductivity due to the
diminishing $\Delta$.

The next three layers ($d$ = 90-150\AA) complete an oscillatory cycle. The
curves are too noisy to discern the more delicate features but the general
behavior is sufficiently clear: the superconductivity is highly suppresed and
the LOFF states develop $d$-dependent coherent oscillations, with smaller and
smaller amplitude. The sign of these oscillations depend on the energy. In
Fig.~\ref{fig:F4}, the reduced DOS, $G(E)/G_0 -1$, oscillates in antiphase at
energy E = 0 with respect to the E = $\pm\Delta$ values. At lower temperatures
the zero energy conductance is lower than the theoretically predicted curve
(Fig.~\ref{fig:F4}a), due to the significant presence of the BTRS induced
ZBCP split. The BTRS states disappear at a temperature lower than BSCCO
$T_c$, so that the 150\AA\ spectrum presents a 60K ZBCP without the split.
Consequently, the points taken at temperatures nearer to $T_c$ approach
better the theoretical expectation.

In summary, we constructed planar Fe/Ag/BSCCO junctions, with the Fe region
imparted in layers of equal thickness. Temperature dependent tunneling spectra
were collected on each layer, probing a reproducible spatial dependent instance
of the LOFF state. The spectral features are affected by the pair-braking
exchange field in the FM region, the thickness of this area and the
unconventional pairing symmetry in the BSCCO. As the Fe thickness is increased,
the conductance spectra show decaying oscillations with energy dependent
amplitude and sign. The ZBCP is split regardless the thickness, but dependent
on the temperature, due to the presence of a subcomponent added to the dominant
d-wave component at the junction interface to form a BTRS state. The spectrum
is asymmetric at low temperature due to the spin-polarization.

This work is supported by NSF Grant No. DMR9972071.


\setlength{\textwidth}{7.1in}
\setlength{\textheight}{9.7in}
\small \rm

\end{document}